\documentclass{article}
\usepackage[english]{babel}
\usepackage[letterpaper,top=2cm,bottom=2cm,left=3cm,right=3cm,marginparwidth=1.75cm]{geometry}

\usepackage{amsmath}
\usepackage{amssymb}
\usepackage{graphicx}
\usepackage{subcaption}
\usepackage{dirtytalk}
\usepackage{tikz}
\usetikzlibrary{math}
\usetikzlibrary{plotmarks}
\usepackage{listings}
\usetikzlibrary{positioning}
\usetikzlibrary{angles, quotes}
\usepackage{float}
\usepackage{listofitems} % for \readlist to create arrays
\usetikzlibrary{arrows.meta} % for arrow size
\usepackage[outline]{contour} % glow around text
\contourlength{1.4pt}
\usepackage{booktabs}
\usepackage{pgfplots}
\usepgflibrary{shadings}
\usetikzlibrary{shadings}
\pgfplotsset{compat=1.18}
\usepackage{xcolor}
\usepackage{graphicx}
\usepackage{multirow}
\usetikzlibrary {matrix}
\usetikzlibrary{positioning,arrows,calc,math,angles,quotes}
\usepackage[colorlinks=true, allcolors=blue]{hyperref}

\usepackage{changes}

\title{Particle Trajectory Prediction in Discrete Element Simulations using a Graph-Based Interaction-Aware Model}

\author{Abhishek Setty\textsuperscript{12$\dagger$}, Lukas Morand\textsuperscript{1}, Poojitha Ramachandra\textsuperscript{1}, Claas Bierwisch\textsuperscript{1}}
\date{}

\usepackage{titling} 
\preauthor{\begin{center}\small}
\postauthor{\par\end{center}}
\vspace{-1em}

\begin{document}

\maketitle

\begin{center}
\small
\textsuperscript{1}Fraunhofer Institute for Mechanics of Materials IWM, Freiburg, Germany \\
\textsuperscript{2}Forschungszentrum Jülich, Institute of Quantum Control (PGI-8), D-52425 Jülich, Germany \\
\textsuperscript{$\dagger$}\texttt{a.setty@fz-juelich.de}
\end{center}

\begin{abstract}
This study explores the applicability of a graph-based interaction-aware trajectory prediction model, originally developed for the transportation domain, to forecast particle trajectories in three-dimensional discrete element simulations. The model and our enhancements are validated at two typical particle simulation use cases: (i) particle flow in a representative unit cell with periodic boundary conditions (PBCs) in combination with sinusoidal velocity profile and (ii) shear flow in a representative unit cell with Lees\texttt{-}Edwards boundary conditions (LEBCs). For the models to learn the particle behavior subjected to these boundary conditions requires additional data transformation and feature engineering, which we introduce. Furthermore, we introduce and compare two novel training procedures for the adapted prediction model, which we call position\texttt{-}centric training (PCT) and velocity\texttt{-}centric training (VCT). The results show that the models developed for the transportation domain can be adapted to learn the behavior of particles in discrete element simulations.
\end{abstract}

\textbf{Keywords:} Trajectory Prediction, Particle Simulation, Discrete Element Method, Periodic Boundary Conditions, Lees\texttt{-}Edwards Boundary Conditions.

%\newpage

\section{Introduction}

\subsection{Motivation}

The study of granular materials and complex fluid flow, which involves particle interactions governed by Newton's equations of motion, holds significant importance for various industrial processes, such as food powder production, pharmaceutical manufacturing, and bulk handling \cite{teunou1999characterisation, salman2006granulation, brown1998silos}. To compute the motion of typically thousands to millions of particles in a system, the discrete element method (DEM) is the means of choice \cite{cundall1979discrete}. By accurately modeling the particle interactions using simple contact forces, DEM simulations can provide insights into the macroscopic properties of the material based on its microscopic interactions \cite{kruggel2007review, thornton2011investigation}. However, DEM simulations face significant constraints related to the time step size and the associated computational effort. The time step must be small enough to accurately capture the dynamics of particle interactions, particularly during collisions \cite{o2004selecting}. This requirement ensures numerical stability and accuracy but also increases the total number of time steps needed, thereby escalating the computational effort. These simulations are computationally intensive due to the necessity of calculating forces and updating positions for a large number of particles at each time step \cite{kruggel2008selection}. Additionally, storing the state of each particle and their interactions demand substantial memory, especially for large-scale simulations. 

To manage the computational load, simulations often need to be parallelized, which introduces further complexity in terms of programming and resource management, especially when using a graphics processing unit (GPU) \cite{dong2022efficient, skorych2022parallel}.
The use of machine learning approaches to predict particle motion has the potential to remove the time-step constraint in DEM simulations, resulting in several significant benefits. Firstly, it allows for larger time steps, thereby reducing the total number of time steps required for a simulation. Also, there is a scope for a significant reduction in computational effort and time, making it possible to simulate larger systems or longer time periods \cite{adamczyk2024machine, kishida2023development}. Additionally, machine learning models are able to capture complex particle interactions in the domains such as gas\texttt{-}particle flow \cite{ouyang2022machine, zhu2020machine}, fluid flow for spherically shaped particles \cite{wan2018machine}, hydrodynamics \cite{zhu2022review, woodward2023physics} and also in detecting contacts of non-spherical particles \cite{lai2022machine}, improving the realism of the simulations. This can enable more detailed and precise predictions of material behavior under various conditions. Furthermore, the reduced computational load frees up resources, allowing for more extensive parameter studies and optimizations, ultimately accelerating research and development.

In general, the works described above are part of a current trend in material science, namely the emergence of data-driven methods \cite{agrawal2016perspective}. Since the last decade, data-driven models have been applied to various problems including representing materials relations \cite{dornheim2024neural}, designing materials and processes \cite{morand2023machine}, parameter identification \cite{morand2024data} or experimentation analysis \cite{durmaz2021deep}. Recent trends are to bring together modern (semantic) data management techniques and software applications (e.g. machine learning applications) on platform environments, such as described in \cite{nahshon2024}. By providing data following the FAIR principles (findable, accessible, interoperable, and reusable), we can expect data-driven methods to reach a new level of effectiveness \cite{scheffler2022fair}.

\subsection{Related Work}
One of the recent examples of accelerating DEM simulations using machine learning is the reduced-order modeling approach described in \cite{wallin2021data}. Therein, a ridge regression model trained on data from numerous DEM simulations is used to predict new system states. Another example is the use of pre-trained Convolutional Neural Networks (CNNs) to replace direct calculations of particle-particle collision and particle-boundary collision behavior and thereby accelerate the computational speed of a DEM simulation $(\text{Conv-DEM})$ \cite{lu2021machine}. This Conv-DEM model has been tested on granular flow in a rotating drum and a hopper. Instead of data-driven methods for predicting granular flow, an image-driven prediction was proposed for a wedge-shaped hopper DEM simulation \cite{liao2021image}. However, this image-based method has the limitation of not being able to capture particle interactions and predict future trajectories. 

One of the early ideas to capture interactions among particles over time is to construct dynamic interaction graphs (aka DPI-Nets) \cite{li2018learning}. This has proven to be working well for interactions of rigid bodies, deformable objects, and fluids. Alternatively, a continuous CNN has been developed and argued to be more accurate and faster than DPI-Nets for learning fluid flows \cite{ummenhofer2019lagrangian}. Inspired by these results in fluid simulations by continuous convolutions, Walters et al. \cite{walters2020trajectory} tackled the problem of trajectory prediction of vehicles using continuous convolutions. In addition to this approach, graph networks have emerged as a promising technique for learning particle simulations, where each particle is represented as a node in a graph \cite{sanchez2020learning}. Similarly, graph-based trajectory prediction is gaining popularity in autonomous driving, which is of particular importance for the presented work. 

One of the popular choices in the transportation domain is the graph-based interaction-aware trajectory prediction (GRIP++) model introduced by Li et al. \cite{li2019grip++}. This method was ranked 1 in the ApolloScape trajectory competition in October 2019. GRIP++ was developed on top of other established methods such as convolutional social pooling \cite{deo2018convolutional}. Among the survey done on trajectory prediction methods for autonomous driving \cite{huang2022survey}, GRIP++ was found to be one of the preferred ones in graph convolutional network-based methods. Despite its success, to the authors' knowledge, the GRIP++ model has not yet been applied to problems outside of the transportation domain. Only in a preliminary study, which formed the basis for the present work, the advantages of using GRIP++ in molecular dynamics were analyzed \cite{Ramachandra2021evolution}. Therefore, together with our preliminary study \cite{Ramachandra2021evolution}, this work is the first to assess the applicability of the GRIP++ model outside of the transportation domain.

\subsection{Contribution}
In this study, we adapt, refine, and assess a trajectory prediction model from the transportation domain, namely GRIP++ \cite{li2019grip++}, to capture the flow of particles in representative unit cells. As particles in such problems behave similarly to cars on a road (i.e. particles move past each other depending on their direct neighbours), we enhance GRIP++ and demonstrate it at two typical particle flow use cases. The main contributions of this work are as follows:
\begin{itemize}
    \item Investigate whether particle trajectories in a DEM simulation can be effectively predicted using the GRIP++ framework.
    \item Modification of the original GRIP++ model to process and learn particle behavior in 3D.
    \item Novel enhanced training strategies for GRIP++, namely position\texttt{-}centric training (PCT) and velocity\texttt{-}centric training (VCT).
    \item Introduction of data preprocessing and transformation steps to enable machine learning models to process data from representative unit cells with simple periodic and Lees\texttt{-}Edwards boundary conditions.
    \item Introduced feature engineering for distinguishing between particles undergoing standard periodic boundary conditions and those subject to Lees-Edwards shifted periodicity. This differentiation is made to extract the information of particles exiting these two boundary conditions. Finally, we developed a VCT-based model to process the updated data after feature engineering, resulting in improved prediction quality in the LEBCs use case.
\end{itemize}

\subsection{Paper Structure}
The structure of the paper is as follows: In Section \ref{sec:methods}, we give a brief introduction to the GRIP++ model as adapted to the 3D particle prediction use case and introduce the data transformation steps needed for simple periodic and Lees\texttt{-}Edwards boundary conditions. In addition, we present the novel particle trajectory prediction methods, namely position\texttt{-}centric training and velocity\texttt{-}centric training. In Section \ref{sec:results}, the performance of the proposed methods on the introduced use cases is shown. The results are discussed in Section \ref{sec:discussion}. Finally, the work is summarized and a future outlook is given in Section \ref{sec:conclusion}.

\section{Methods}\label{sec:methods}
\subsection{Modified GRIP++ Model}
\label{sec:model}
\begin{figure}[H]
    \centering
    \includegraphics[width=0.8\textwidth]{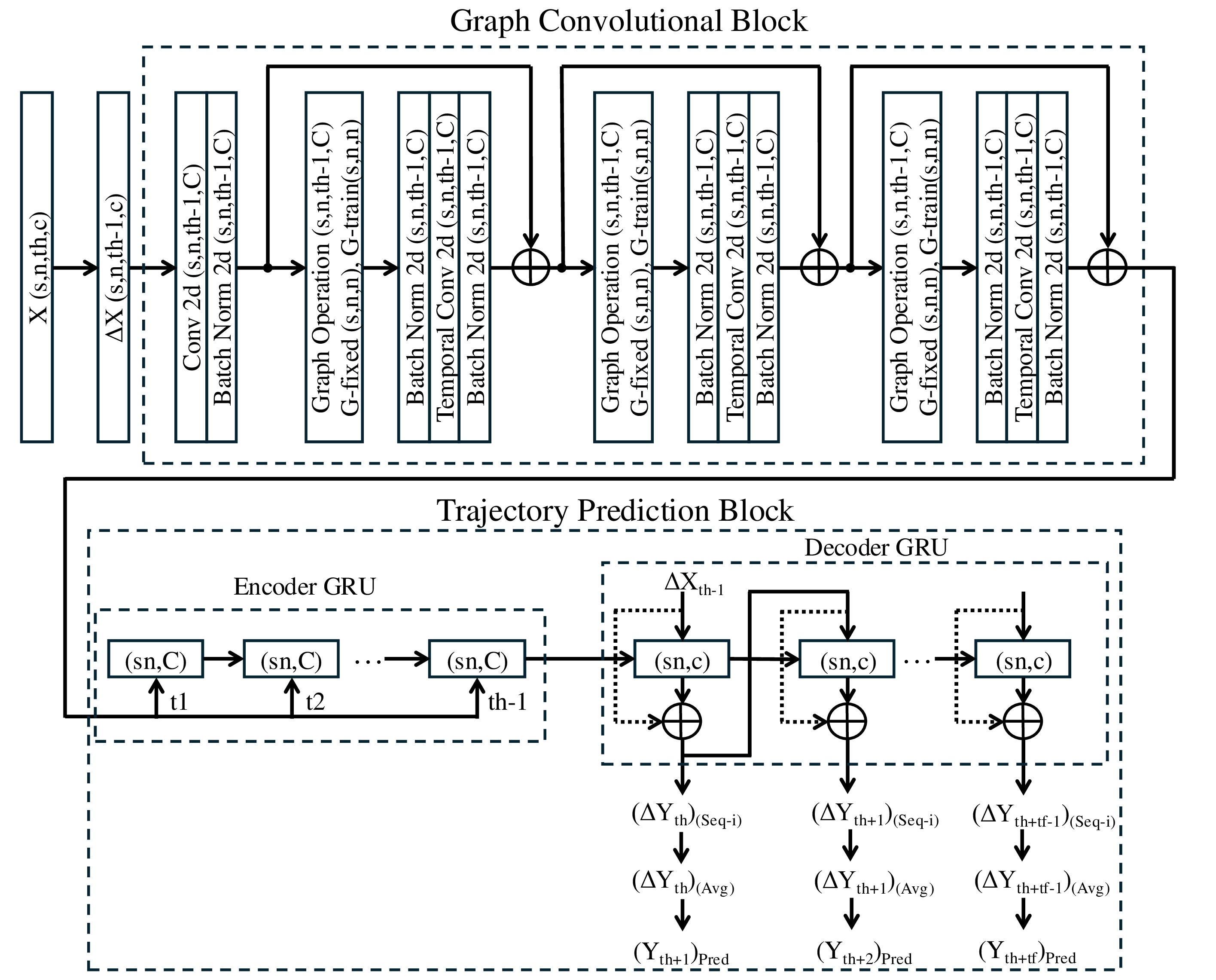}
    \captionsetup{width=\textwidth}
    \caption{Illustration of the information flow in the GRIP++ model as adapted to the $c$-dimensional case.}
    \label{fig:9_GRIP++}
\end{figure}

The GRIP++ model is adapted to predict the trajectory of particles. This method works by employing a convolution operation on graph-based data. Each particle in the DEM simulation is treated as a node to form a graph, similar to the graph network simulator method described in \cite{sanchez2020learning}. The interaction among a node and its surrounding nodes is represented by an edge in the graph. This constitutes a spatial interaction among nodes in a graph. A node traveled over time steps\footnote{Note that in the following, with time steps, we do not refer to DEM increments but to equidistant discrete points in simulation time.} is represented by a sequence of edges forming a trajectory over history. This constitutes a temporal sequence of a node in the graph. The core idea is to learn a mapping function that extracts significant information from these edges and predicts the future trajectory. The GRIP++ model consists of a data preprocessing for input representation followed by a graph convolutional block for extracting the spatial and temporal features and a trajectory prediction block containing several sequence-to-sequence networks each containing an encoder-decoder-based gated recurrent units \cite{cho2014learning}. An overview of the components of the modified GRIP++ model is illustrated in Figure \ref{fig:9_GRIP++}.

\subsubsection{Input Representation}
The original GRIP++ model for the transportation domain is designed for processing inputs in 2D Cartesian space. In this work, we consider the coordinates of particles in 3D Cartesian space. 
Therefore, we formulate the model for $c$ dimensional space. Consider the input data $X$ of the model which consists of $s$ simulations where each simulation is a trajectory history over $t_h$ time steps of all particles, given as

\begin{equation}
    X = \left[ \begin{array}{ccc}
    p^{(11)} & \hdots & p^{(1t_h)} \\
    \vdots & \ddots & \vdots \\
    p^{(s1)} & \hdots & p^{(st_h)} \end{array} \right]
\end{equation}
where
\begin{equation}
    p^{(it)} = \left[ \begin{array}{ccc}
    x_{11}^{(it)} & \hdots & x_{1n}^{(it)} \\
    \vdots & \ddots & \vdots \\
    x_{c1}^{(it)} & \hdots & x_{cn}^{(it)} \end{array} \right]
\end{equation}
are the coordinates in $c$ dimensions at time $t$ of all $n$ particles in $i^{th}$ simulation. This information can be represented in an array of the shape $(s, n, t_h, c)$. In \cite{li2019grip++}, it is proposed to learn velocities $(p^{t+1} - p^{t})$ rather than positions of trajectory history. Please note that the velocity here refers to the relative distance vector and the division by the time step (which is a constant) is ignored. Therefore, the input representation of the data is given by
\begin{equation}
    \Delta X = \left[ \begin{array}{ccc}
    \Delta p^{(11)} & \hdots & \Delta p^{(1(t_h-1))} \\
    \vdots & \ddots & \vdots \\
    \Delta p^{(s1)} & \hdots & \Delta p^{(s(t_h-1))} \end{array} \right]
\end{equation}
The output of the model is the future positions of all the particles in all the simulations from time step $t_h + 1$ to $t_h + t_f$, such that,
\begin{equation}
    Y = \left[ \begin{array}{ccc}
    p^{(1(t_h+1))} & \hdots & p^{(1(t_h+t_f))} \\
    \vdots & \ddots & \vdots \\
    p^{(s(t_h+1))} & \hdots & p^{(s(t_h+t_f))} \end{array} \right]
\end{equation}
where $t_f$ is the predicted end time step.

\subsubsection{Graph Construction} \label{sec:graph_construction}
The motion of a particle is impacted by the movement of its surrounding particles. Therefore, the inter-particle interaction is captured using a graph $G=\{V, E\}$, where each particle is represented by a node in the set $V$ and an edge in the set $E$. Node set $V$ is given as
\begin{equation}
V = \{v_{i}|i = 1, \dots, n, t = 1, \dots, t_h\},
\end{equation}
where $n$ is the number of particles and $t_h$ is the observed time steps. The feature vector $v_{it}$ on a graph node is the coordinate of the $i$th particle at time $t$. In this work, the interaction among particles is represented by an edge set $E$ at the last observed time step $t_h$, such as $E = \{v_{it_h}v_{jt_h}|(i,j\in D)\}$, where $D$ is a threshold radius within which two particles are interacting with each other. The distance between two particles is calculated between the centers of the sphere-shaped particles and we chose $D$ to be three times the radius of a particle. This graph construction is modified by an adjacency matrix $A$, such as,
\begin{equation}
    A[i, j] = 
    \begin{cases}
    1, &\text{if edge} \; \langle v_{it_h}, v_{jt_h}\rangle \in E \\
    0, &\text{otherwise}
    \end{cases}
\end{equation}

The matrix $A$ has a size of $(n, n)$. Based on the graph constructed at the last observed time step $t_h$, the model is trained to predict the desired time steps at once. This graph construction is fixed and does not change during the training phase. The matrix $A$ is used to construct the fixed graph in Section \ref{sec:graph_convolution}.

\subsubsection{Graph Convolutional Block} \label{sec:graph_convolution}
Once the input data is preprocessed $\Delta X := \mathbb{R}^{s \times n \times t_h-1 \times c}$, it is passed through a first 2D convolutional layer with $(1 \times 1)$ kernel size to increase the number of channels. It maps $c$\texttt{-}dimensional data into higher dimensional space $C$ such that the output has a shape of $(s, n, t_h-1, C)$, where $C$ is the updated number of channels. In this work, we chose $C=64$.

After convolution, the data is fed into the consecutive graph operation and temporal convolutional layers, which are repeated three times (refer Figure \ref{fig:9_GRIP++}). The graph operation layer captures the features in the spatial dimension, and the temporal convolutional layer processes them through temporal space. Skip connections are added between these layers to ensure that larger gradients are propagated to the initial layers to tackle the problem of vanishing gradients. The initial layers can also learn as fast as the final layers. Batch normalization layers are employed to improve the training stability of the model. 

The graph operation layer consists of two graphs: (\romannumeral 1) a fixed graph $G_{fixed}$ constructed based on the last observed time step, and (\romannumeral 2) a trainable graph $G_{train}$ which captures the dynamic interactions that have trainable parameters of shape $(s, n, n)$ which is same as the fixed graph. The fixed graph is normalized to make sure the values are in the same range after the graph operation layer, such that
\begin{equation}
    G_{fixed} = \Lambda^{-\frac{1}{2}} A \Lambda^{-\frac{1}{2}},
\end{equation}
where matrix $A$ is defined in the Section \ref{sec:graph_construction} and $\Lambda$ is given as:
\begin{equation}
    \Lambda^{ii} = \sum_k (A^{ik}) + \alpha,
\end{equation}
$\alpha$ is set to $0.001$ to avoid empty rows in $A$. 

Consider the output of the 2D convolutional layer is given by $f_{conv}$, then it is fed into the graph operation layer $f_{graph}$, such that,
\begin{equation}
     f_{graph} = (G_{fixed} + G_{train}) f_{conv}.
\end{equation}
The shape of the $f_{graph}$ is $(s, n, t_h-1, C)$ which is the same as $f_{conv}$. The output of the graph operation layer is fed into the temporal convolutional layer in which the kernel size is of $(1 \times 3)$ to process the data along the temporal dimension. Within the temporal convolution layer, the stride is set to $1$, and appropriate padding is assigned to maintain the output shape to be $(s, n, t_h-1, C)$.

\subsubsection{Trajectory Prediction Block}
\label{sec:model_trajectory_prediction}
The trajectory prediction block consists of several sequence-to-sequence-based (Seq2Seq) \cite{sutskever2014sequence} encoder-decoder GRU (Gater Recurrent Units) networks. The results from all the Seq2Seq networks are averaged in the end. In a single Seq2Seq network, the output of the graph convolutional block is fed into the encoder GRU network such that each GRU cell out of $t_h - 1$ tokens (a unit cell for a single time step in a sequence network) takes the input of $C$ features and $sn$ samples (number of simulations times the number of particles). The hidden features of the encoder network as well as the velocity values of all the particles at the last time step are fed into the decoder GRU network to predict the velocities at the current time step. 

These predicted velocities are added to the previous position coordinates to get the position coordinates at the current time step. The output of the current time step is fed as the input into the next GRU cell. In addition, residual connections are added between the input and the output of each cell in the decoder GRU network to force the model to learn the change in the velocities. In this work, we optimized the model hyperparameters (kernel size, channels, threshold distance, number of GRU layers, features in hidden state, number of epochs) manually until we obtained satisfactory results. For the architecture, we chose two Seq2Seq networks, where each network has two GRU layers, each of $350$ features in hidden state. We trained the model for 6000 epochs using the mean squared error (MSE) loss function and Adam optimizer with default parameters in PyTorch \cite{paszke2019pytorch}.

\subsection{Boundary Conditions}\label{sec:bc}

For performing particle simulation in a representative unit cell, the definition of boundary conditions is essential. In this work, two different boundary conditions, namely, periodic boundary condition (PBC) and Lees\texttt{-}Edwards \cite{lees1972computer} boundary conditions (LEBC) are considered. To apply the GRIP++ model to the prediction of particle flows in a representative unit cell, the boundary conditions of the unit cell have to be integrated into the model. We do this by a data transformation step to the model inputs and outputs. In the following, we describe the data transformation rules for periodic and Lees\texttt{-}Edwards boundary conditions.

\subsubsection{Periodic Boundary Conditions With Sinusoidal Velocity Profile}
\label{sec:pbc}
The purpose of PBCs in the context of DEM simulations is to create a more realistic and efficient representation of an infinite or large system by eliminating edge effects and reducing computational costs. By replicating the simulation box periodically in all directions, particles that move out of one side of the box re-enter from the opposite side, maintaining a continuous and seamless environment \cite{allen1987computer}. This approach is particularly useful for several reasons. It allows for the study of bulk properties of materials without the interference of boundary effects that can distort the results. Additionally, PBCs enable the simulation of smaller representative volumes, which significantly reduces the computational resources required compared to simulating a much larger system. This makes it feasible to investigate the behavior of materials and systems over longer time scales and larger spatial scales while maintaining accuracy and efficiency.

\begin{figure}[H]
    \centering
    \includegraphics[width=0.8\textwidth]{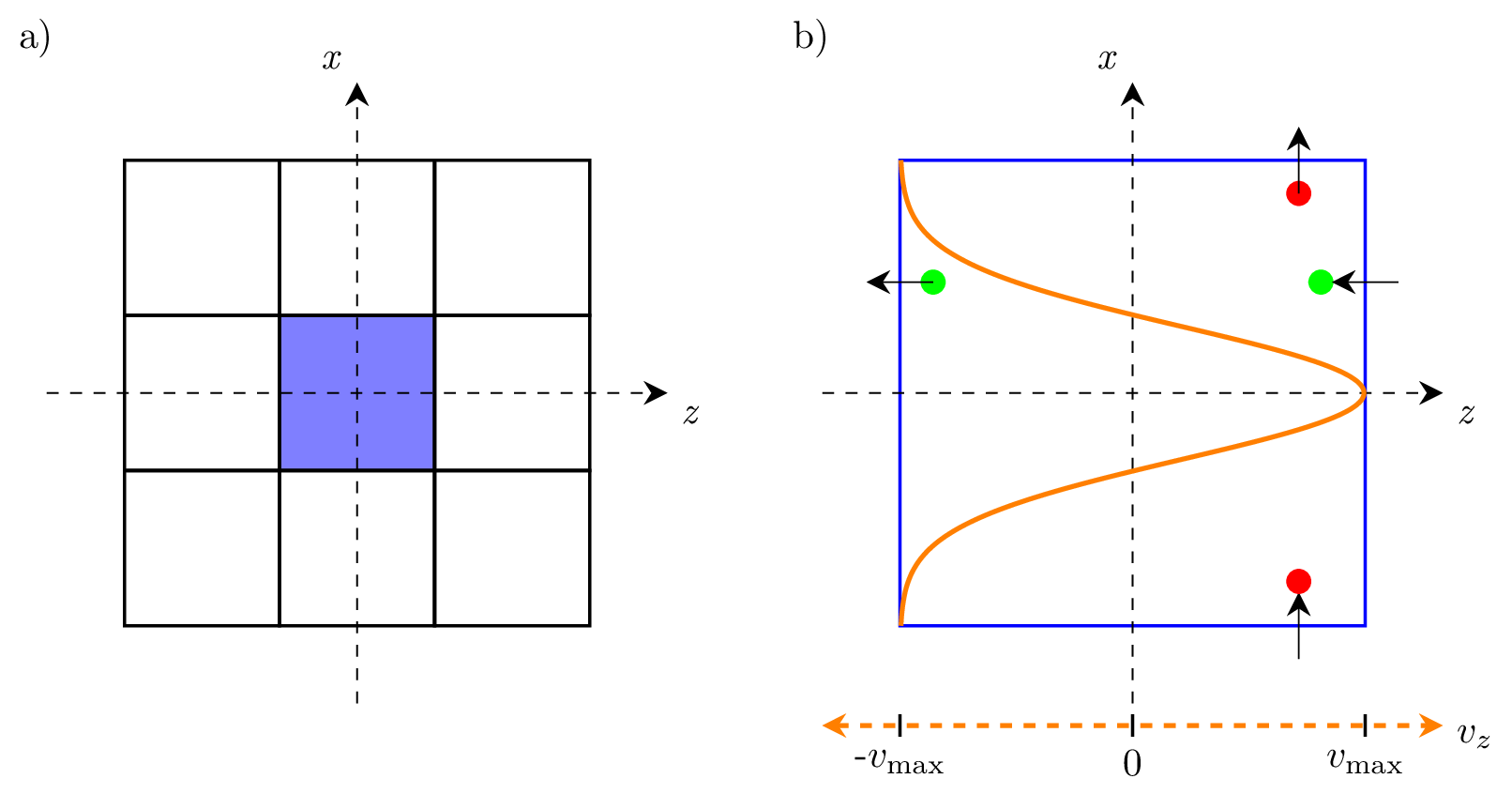}
    \captionsetup{width=\textwidth}
    \caption{a) PBCs in $x$ and $z$ axes. b) The simulation cell with two example particles (red and green) exiting and entering the $x$ and $z$ boundaries, respectively. In addition, the velocity profile (orange) of all particles in $z$-direction is represented with respect to $x$-axis.}
    \label{fig:1_Periodic}
\end{figure}

Figure \ref{fig:1_Periodic}a illustrates the characteristics of PBCs. The colored cell represents the simulated cell and the surrounding cells are the image cells. For simplicity of representation, the simulation cell is represented in two dimensions using $x$ and $z$ axes. Each particle has Cartesian coordinates in three dimensions with the origin being the center of the cell. Consider the box length in three dimensions given by $(l_x, l_y, l_z)$. In addition, Figure \ref{fig:1_Periodic}b illustrates a sinusoidal velocity profile in a representative unit cell, and two example particles are shown while exiting the $x$ and $z$ boundaries. A typical velocity profile is created by a periodic acceleration $a_z(x) = a \cos(2 \pi x / l_x)$ acting on the $z$-component of the particles' velocities. Please note that the term $x$ boundary in a 3D simulation cell means $y$-$z$ plane at $x=l_x/2$ or $x=-l_x/2$ depending on where the particle exited or entered. Similarly, $y$ and $z$ boundaries wherever mentioned are considered accordingly.

The data preprocessing for the GRIP++ model requires computing a fixed graph which consists of computing the distance between every particle with every other particle at the current time step. With PBCs, the distance between any two particles is computed using the minimum-image-convention. This essentially means the minimum distance between those two particles among all the pairs with the image particles. If the relative distance between two particles at a particular time step $t$ is given as
\begin{equation}
    (r_x, r_y, r_z) = (x_a - x_b, y_a - y_b, z_a - z_b),
\end{equation}
then the relative distance is updated based on conditional statements as follows,
\begin{equation}\label{eq:r_p}
\begin{aligned}
\text{if} \; &r_x < -\frac{l_x}{2}, \; &\text{then} \; (r_x)_{\text{update}} = r_x + l_x,\\
\text{if} \; &r_x \geq \frac{l_x}{2}, \; &\text{then} \; (r_x)_{\text{update}} = r_x - l_x,\\
\text{if} \; &r_y < -\frac{l_y}{2}, \; &\text{then} \; (r_y)_{\text{update}} = r_y + l_y,\\
\text{if} \; &r_y \geq \frac{l_y}{2}, \; &\text{then} \; (r_y)_{\text{update}} = r_y - l_y,\\
\text{if} \; &r_z < -\frac{l_z}{2}, \; &\text{then} \; (r_z)_{\text{update}} = r_z + l_z,\\
\text{if} \; &r_z \geq \frac{l_z}{2}, \; &\text{then} \; (r_z)_{\text{update}} = r_z - l_z.\\
\end{aligned}
\end{equation}
Note that all the particles considered in this work are spherical with the radius of the particle being mentioned in Table \ref{tab:Variables}. While the distance between two particles is calculated in Cartesian coordinates as described above, we propose transforming this periodic dataset from Cartesian coordinates to $\sin$ and $\cos$ values \cite{Ramachandra2021evolution}. This is advantageous for the GRIP++ model as it avoids jumps in the particle trajectories when particles exit and reenter the unit cell. Consider a particle location in Cartesian coordinates as $(x, y, z)$, then the periodic representation of this particle includes six values, which are defined as,
\begin{equation} \label{eq:sin_cos}
\begin{aligned}
    x_{\sin} &= \sin{\left(\frac{2\pi x}{l_x}\right)}, & y_{\sin} &= \sin{\left(\frac{2\pi y}{l_y}\right)}, & z_{\sin} &= \sin{\left(\frac{2\pi z}{l_z}\right)},\\
    x_{\cos} &= \cos{\left(\frac{2\pi x}{l_x}\right)}, & y_{\cos} &= \cos{\left(\frac{2\pi y}{l_y}\right)}, & z_{\cos} &= \cos{\left(\frac{2\pi z}{l_z}\right)}.
\end{aligned}
\end{equation}
The input of the GRIP++ model is the velocities of all particles which are the differences of position coordinates between equidistant time steps. As in this study, we consider equidistant time steps, we ignore the time interval and directly use the relative distance between the two time steps as a velocity vector. For example, consider the $\Delta x_{\sin}$ value at time $t$ as $\Delta x_{\sin}^t = x_{\sin}^{t+1} - x_{\sin}^t = \sin{(\frac{2 \pi x^{t+1}}{l_x})} - \sin{(\frac{2 \pi x^t}{l_x})}$. To transform the new data representation back to Cartesian coordinates, we need to invert the relation described in Equation \ref{eq:sin_cos}:
\begin{equation}
    x = \frac{l_x}{2 \pi} \arctan\left({\frac{x_{\sin}}{x_{\cos}}}\right), \hspace{0.5cm}
    y = \frac{l_y}{2 \pi} \arctan\left({\frac{y_{\sin}}{y_{\cos}}}\right), \hspace{0.5cm}
    z = \frac{l_z}{2 \pi} \arctan\left({\frac{z_{\sin}}{z_{\cos}}}\right).
\end{equation}

\subsubsection{Lees\texttt{-}Edwards Boundary Conditions}\label{sec:LEBS}
\begin{figure}[H]
    \centering
    \includegraphics[width=0.8\textwidth]{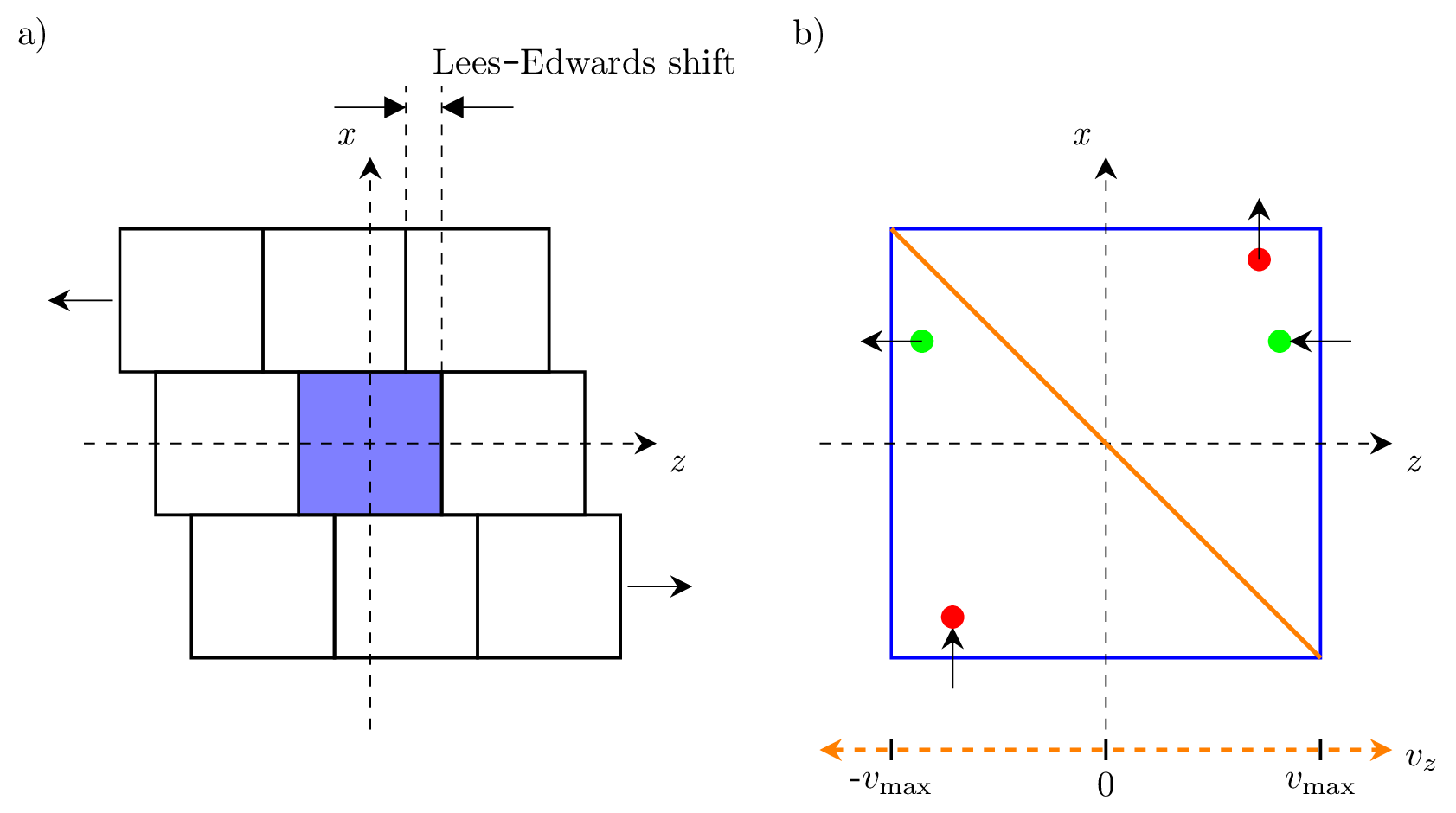}
    \captionsetup{width=\textwidth}
    \caption{a) The LEBCs and the corresponding Lees\texttt{-}Edwards shift. b) Representative unit cell with two example particles (red and green) exiting and reentering the $x$ and $z$ boundaries, respectively. In addition, the velocity profile (orange) of all particles in $z$-direction is represented with respect to $x$-axis.}
    \label{fig:2_Lees_Edwards}
\end{figure}

LEBCs are a specialized type of PBCs used in DEM simulations to model shear flow \cite{lees1972computer}. These conditions are particularly valuable for simulating systems subjected to shear deformation. In this approach, the representative unit cell is periodically replicated, but with a relative velocity applied to the Lees\texttt{-}Edwards boundaries (top and bottom in Figure \ref{fig:2_Lees_Edwards}). This setup creates a shear flow across the system, allowing particles to exit one side of the simulation cell and reenter from the opposite side with an appropriate positional shift to maintain the shear profile. These boundary conditions are beneficial because they enable the controlled simulation of shear flow, which is essential for studying the rheological properties of materials. By using periodic boundaries, edge effects are eliminated that could otherwise distort the simulation results. Additionally, they ensure the continuity of the shear profile across the boundaries, providing a more realistic representation of an infinite system under shear. Overall, LEBCs are a powerful tool for investigating the behavior of materials under shear deformation in DEM simulations.

The difference in this case compared to the simple PBC (refer to Section \ref{sec:pbc}) is the shift in the image cells in the direction of shear flow. The magnitude of the shift is determined by the time frame and the shear rate. An illustration of this process including two particles exiting and reentering the simulation cell is represented in Figure \ref{fig:2_Lees_Edwards}. Note that no acceleration is applied in the case of the Lees\texttt{-}Edwards boundary conditions.

In Figure \ref{fig:2_Lees_Edwards}b, the red particle exits from the top right corner of the box. Therefore, it reenters the unit cell with a shift in $z$-direction. The Lees\texttt{-}Edwards shift (referred as $z_{\text{shift}}$) is calculated for each time frame as follows:
\begin{equation}\label{eq:LES}
\begin{split}
    z_{\text{total-shift}} &= t  k  l_x, \\
    z_{\text{shift}} &= z_{\text{total-shift}} - (z_{\text{total-shift}} \bmod(l_z)) ~ l_z, 
\end{split}
\end{equation}
where $t$ denotes time and $k$ denotes the shear rate applied via the LEBCs. The relative distance between two particles at a particular time step $t$ is calculated similarly to Equation (\ref{eq:r_p}), however, slightly modified due to the Lees\texttt{-}Edwards shift as follows:
\begin{equation}\label{eq:r_le}
\begin{aligned}
\text{if} \; &r_x < -\frac{l_x}{2}, \; &\text{then} \; (r_x)_{\text{update}} = r_x + l_x \; &\text{and} \; (r_z)_{\text{update}} = r_z - z_\text{shift},\\
\text{if} \; &r_x \geq \frac{l_x}{2}, \; &\text{then} \; (r_x)_{\text{update}} = r_x - l_x \; &\text{and} \; (r_z)_{\text{update}} = r_z + z_\text{shift},\\
\text{if} \; &r_y < -\frac{l_y}{2}, \; &\text{then} \; (r_y)_{\text{update}} = r_y + l_y,\\
\text{if} \; &r_y \geq \frac{l_y}{2}, \; &\text{then} \; (r_y)_{\text{update}} = r_y - l_y,\\
\text{if} \; &r_z < -\frac{l_z}{2}, \; &\text{then} \; (r_z)_{\text{update}} = r_z + l_z,\\
\text{if} \; &r_z \geq \frac{l_z}{2}, \; &\text{then} \; (r_z)_{\text{update}} = r_z - l_z.
\end{aligned}
\end{equation}

\subsection{Enhancements of the GRIP++ Model}

In the following, we introduce two types of training procedures that we used to enhance the modified GRIP++ model for improved particle trajectory predictions. We call these position\texttt{-}centric training and velocity\texttt{-}centric training. These training methods are modifications mainly in the decoder section of the trajectory prediction block to improve prediction performance. In the original GRIP++ model, the decoder takes velocity as the input and predicts the velocity at the next time step. The residual connection is used to learn the change in the velocity. However, this method can lead to large error accumulation in predictions over multiple time steps. With the novel training procedures, we want to overcome this drawback.

\subsubsection{Position-Centric Training (PCT)}
\begin{figure}[H]
    \centering
    \includegraphics[width=0.8\textwidth]{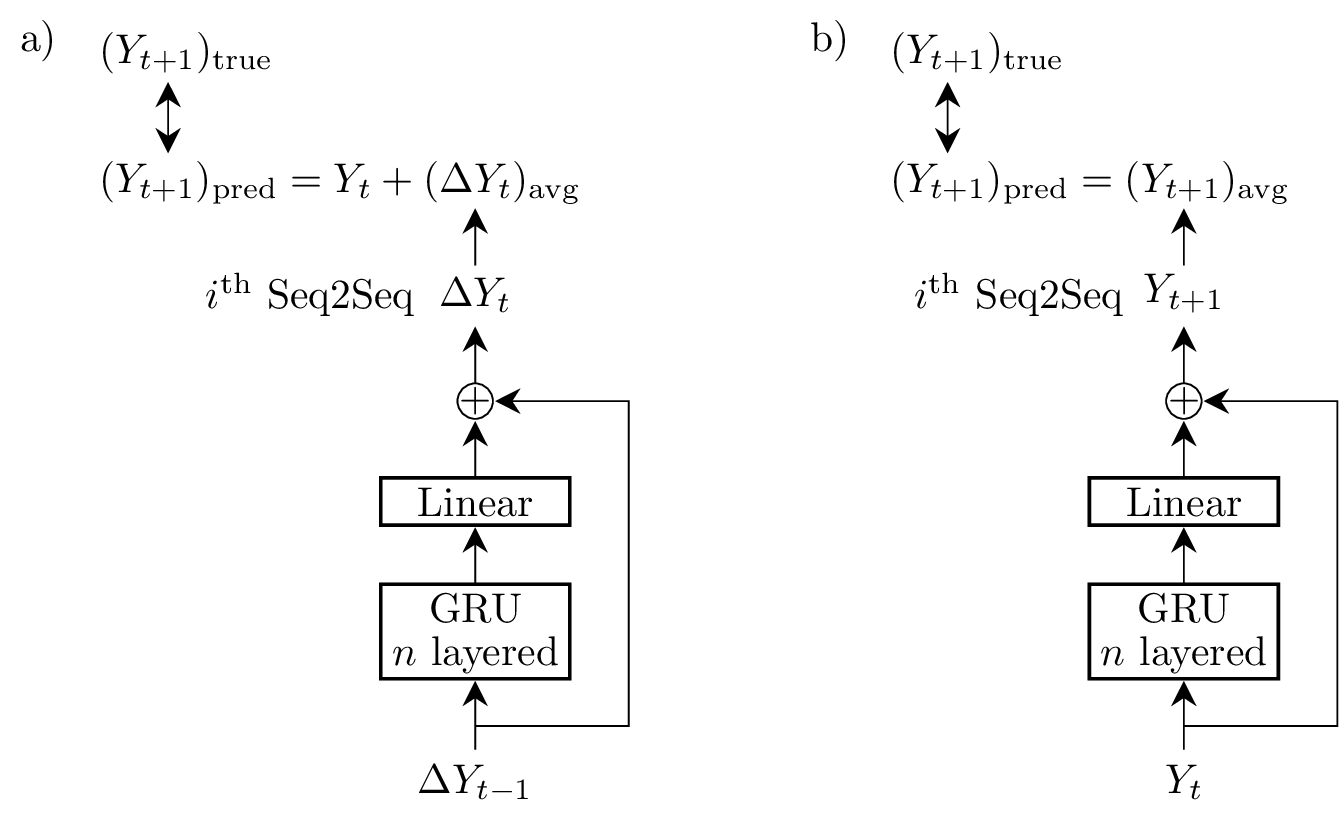}
    \captionsetup{width=\textwidth}
    \caption{a) Decoder section of the encoder-decoder GRU network as used in the modified GRIP++ approach and b) PCT-based model.}
    \label{fig:3_Position_centric_training}
\end{figure}

Consider the decoder section logic of the trajectory prediction block described in Section \ref{sec:model_trajectory_prediction}. We modify the decoder block of the GRIP++ model to consider the position of the particles and call it position\texttt{-}centric training (PCT) (refer to Figure \ref{fig:3_Position_centric_training}). For the sake of simplicity, we represent the multilayered GRU network of the decoder sequence in a concise form. In the figure, the stacked GRU cells are shown as one block for the prediction of a single time step.
The loss criterion is chosen to be a MSE loss and it is calculated between $(Y_{t+1})_{\text{true}}$ and $(Y_{t+1})_{\text{pred}}$ (indicated by the double arrows in Figure \ref{fig:3_Position_centric_training}). This loss function is minimized using the Adam optimizer with default parameters in PyTorch.

\subsubsection{Velocity-Centric Training (VCT) With Feature Engineering}
Only particles that are exiting and reentering through the Lees\texttt{-}Edwards boundaries (refer Figure \ref{fig:2_Lees_Edwards}) experience the Lees\texttt{-}Edwards shift. The remaining particles do not experience a shift and behave following the simple PBCs. This inhomogeneity in the Lees\texttt{-}Edwards use case is causing difficulties for the position\texttt{-}centric training method to learn precisely. To improve the prediction performance for these inhomogeneities, we introduce feature engineering focusing on the velocity of the particles based on their position incorporating the Lees\texttt{-}Edwards shift.

\begin{figure}
    \centering
    \includegraphics[width=0.8\textwidth]{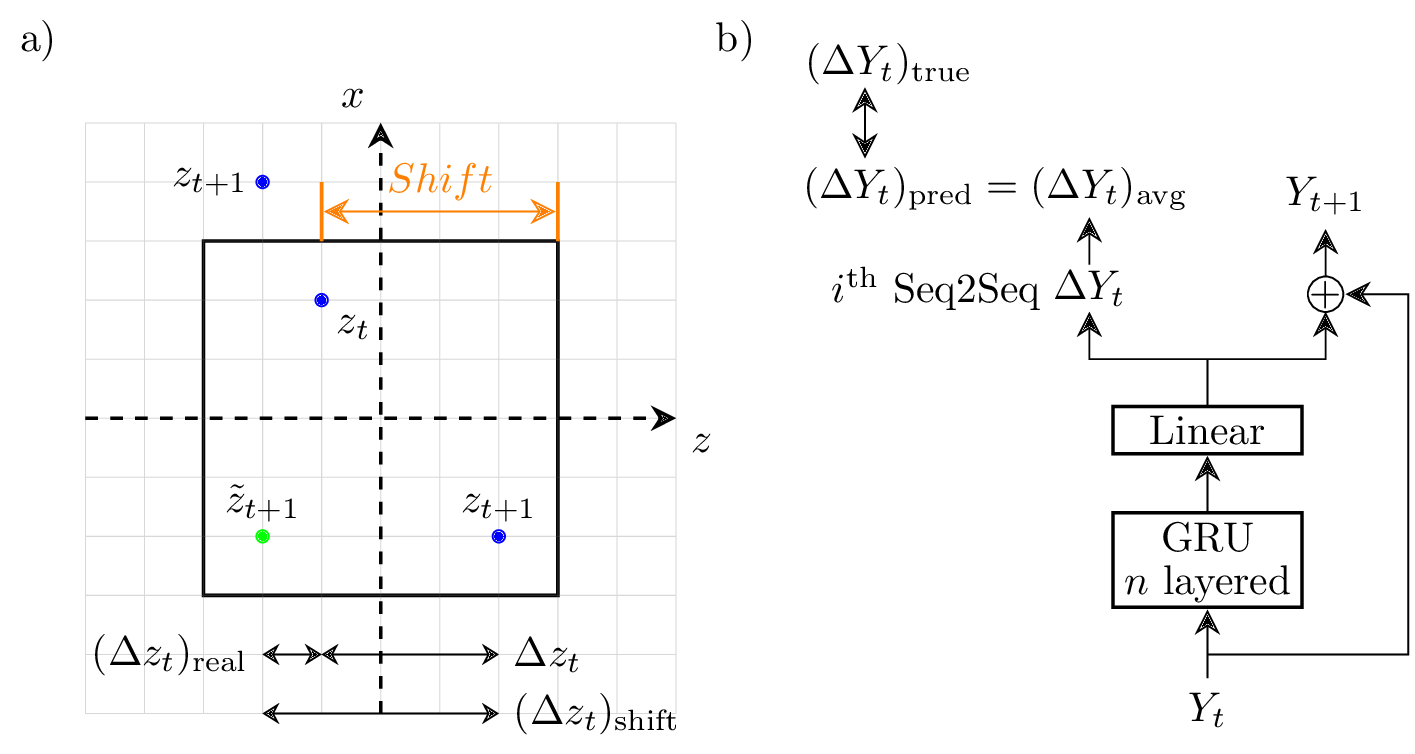}
    \captionsetup{width=\textwidth}
    \caption{a) Example of a particle exiting the $x$-boundary and having a shift in the $z$-direction. b) Decoder section logic of VCT.}
    \label{fig:4_Velocity_centric_traning}
\end{figure}

Consider an example of a particle exiting the Lees\texttt{-}Edwards boundary ($+x$) at a time step $t$ as shown in Figure \ref{fig:4_Velocity_centric_traning}a. This particle reenters the simulation cell with a Lees\texttt{-}Edwards shift (refer as $z_\text{shift}$) in the $z-$direction. Therefore, the $z$ position of that particle at time steps $t$ and $t+1$ will be $z_t$ and $z_{t+1}$, respectively. Concerning the former method of position\texttt{-}centric training, the velocity of the particle is calculated as $\Delta z_t$. However, this is not the real velocity vector as it should consider the image particle at $t+1$ closest to the previous step. Therefore, in this section, we consider the real velocity of the particle $(\Delta z_t)_{\text{real}}$ and include the shift $(\Delta z_t)_{\text{shift}}$ as an additional feature for $z$ positions. In terms of $\sin$ space, this yields
\begin{equation}
\begin{aligned}
    (\Delta z^t_{\sin})_{\text{real}} &= \sin{(\frac{2\pi \tilde{z}_{t+1}}{l_z})} - \sin{(\frac{2\pi z_{t}}{l_z})}, \\
    (\Delta z^t_{\sin})_{\text{shift}} &= \sin{(\frac{2\pi z_{t+1}}{l_z})} - \sin{(\frac{2\pi \tilde{z}_{t+1}}{l_z})}. 
\end{aligned}
\end{equation}
For those particles that are exiting the boundaries ($y$ and $z$ boundaries) other than Lees\texttt{-}Edwards boundary ($x$ boundary), $(\Delta z_t)_{\text{shift}} = 0$. With this additional feature, we differentiate the particles that cross Lees\texttt{-}Edwards boundaries from others. This is the feature engineering we propose for differentiating discontinuous data from continuous data. Therefore, the input data representation has eight features.

To process the updated data after feature engineering, we introduce a velocity\texttt{-}centric training (VCT) method as shown in Figure \ref{fig:4_Velocity_centric_traning}b. The residual connection is used to update the position. The GRU and linear blocks are used to learn the velocities of the particles based on their position. Specifically, we learn the real velocities and the Lees\texttt{-}Edwards shifts of the particles instead of direct locations. This allows the model to recognize and differentiate particles that cross Lees\texttt{-}Edwards boundaries from particles that cross simple periodic boundaries. The calculation of the next position $Y_{t+1}$ accounts for both real velocity and shift in the $z$ axis. For example, $z_{\sin}^{t+1} = z_{\sin}^{t} + (\Delta z_{\sin}^t)_{\text{real}} + (\Delta z_{\sin}^t)_{\text{shift}}$. The loss function is chosen to be MSE and it is calculated between $(\Delta Y_{t})_{\text{true}}$ and $(\Delta Y_{t})_{\text{pred}}$ (indicated by the double arrows in Figure \ref{fig:4_Velocity_centric_traning}b). The model was trained using the Adam optimizer with default parameters in PyTorch.

\subsection{Performance Evaluation}
\begin{figure}[H]
    \centering
    \includegraphics[width=0.99\textwidth]{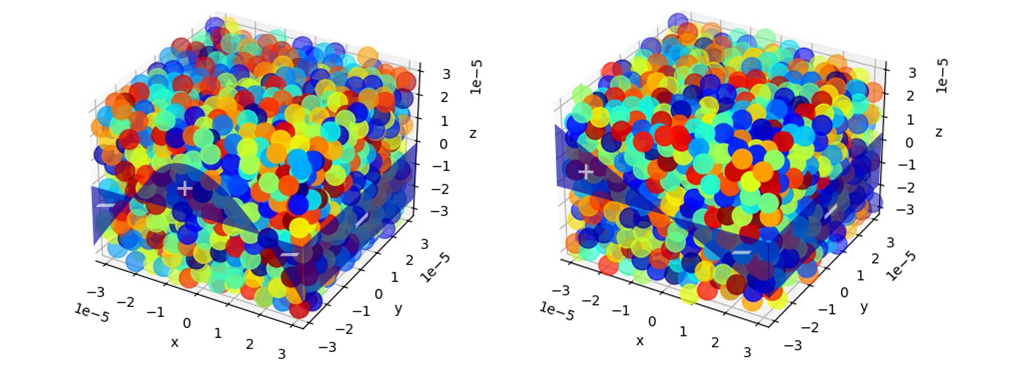}
    \caption{Screenshots of the simulated unit cells including a schematic illustration of the applied velocity profile (shown as a graph with blue shaded regions) with positive and negative markings: Unit cell with simple periodic boundary conditions (left) and unit cell with Less-Edwards boundary conditions (right). The color code is used to better visualize the different particles.}
    \label{fig:simulation_screenshots}
\end{figure}

Two simulation datasets are used for evaluating the model performance for the different boundary conditions described in Section \ref{sec:bc}.  Each dataset originates from one simulation of 2000 particles in motion for 201 equidistant time frames. The interaction of the particles in the simulation are governed by friction, lubrication, and van der Waals cohesion. Figure \ref{fig:simulation_screenshots} shows a screenshot of the simulated unit cells. All the relevant variables used in this work and the model hyperparameters are mentioned in Table \ref{tab:Variables}. 

\begin{table}[ht]
	\centering
	\begin{tabular}{ |c|c|c| } 
\hline
Variable & Usage & Value \\
\hline
$s$ & Simulations & 1 \\ 
$n$ & Particles & 2000 \\
$l_x=l_y=l_z$ & Box length & $3.2 \times 10^{-5}$ \\
$t_h + t_f$ & Total timesteps & 201 \\
$c$ in GRIP++ & Input and output features & 6 \\
$c$ in PCT & Input and output features & 6 \\
$c$ in VCT & Input and output features & 8 \\
$\alpha$ & Parameter in $G_{fixed}$ & 0.001 \\
& Radius of particle & $2.5 \times 10^{-6}$ \\
$D$ & Threshold distance & $3 \times$Radius of particle\\
$k$ & Shear rate in LEBC & 1000 \\
& Kernel in first convolution & $(1 \times 1)$ in $(n \times t)$\\
& Kernel in temporal convolution & $(1 \times 3)$ in $(n \times t)$\\
$C$ & Channels in convolution & 64 \\
& Number of Seq2Seq networks & 2 \\
& Number of GRU layers & 2 \\
& Hidden features in GRU & 350 \\
& Training epochs & 6000 \\
\hline
\end{tabular}
	\caption{Overview of variables and hyperparameters.}
	\label{tab:Variables}
\end{table}

After training the model, the accuracy of the predicted results is evaluated using the root mean squared error (RMSE) measure. To evaluate RMSE, consider our output data in true and predicted values of shape $(\text{simulations, particles, time steps, coordinates})$ as $(s, n, t, c)$. Firstly, the relative distance between the true and predicted position is calculated using Equations (\ref{eq:r_p}) and (\ref{eq:r_le}) depending on the use case. It is calculated as $r_c$, where $c \in \{x, y, z\}$ for all $s$ simulations, $n$ particles, and $t$ time steps. Then the RMSE is evaluated as, 
\begin{equation}\label{eq:rmse}
    \text{RMSE} = \sqrt{\frac{\sum_{s,n,t,c}r^2_{sntc}}{snt}}.
\end{equation}
Note that in the above Equation \ref{eq:rmse}, the denominator $snt$ represents the product of $s$, $n$, and $t$.

\section{Results}
\label{sec:results}
To compare the performance of our modified GRIP++ model and the proposed enhancements, all studies are conducted with a fixed set of model hyperparameters (as they are mentioned in Section \ref{sec:model}). In our study, the evaluation datasets are split into input and output time steps so that the model has to infer future particle trajectories (output time steps) from given past trajectory data (input time steps). We have studied various time step splits $((t_h-1)\textendash tf)$ such as 190-10, 180-20, 150-50, and 100-100 and compared the performance in terms of RMSE score (refer to Equation \ref{eq:rmse}).

\subsection{Evaluation on Simple PBCs}
\label{sec:results_periodic}
\begin{figure}[H]
    \centering
    \includegraphics[width=\textwidth]{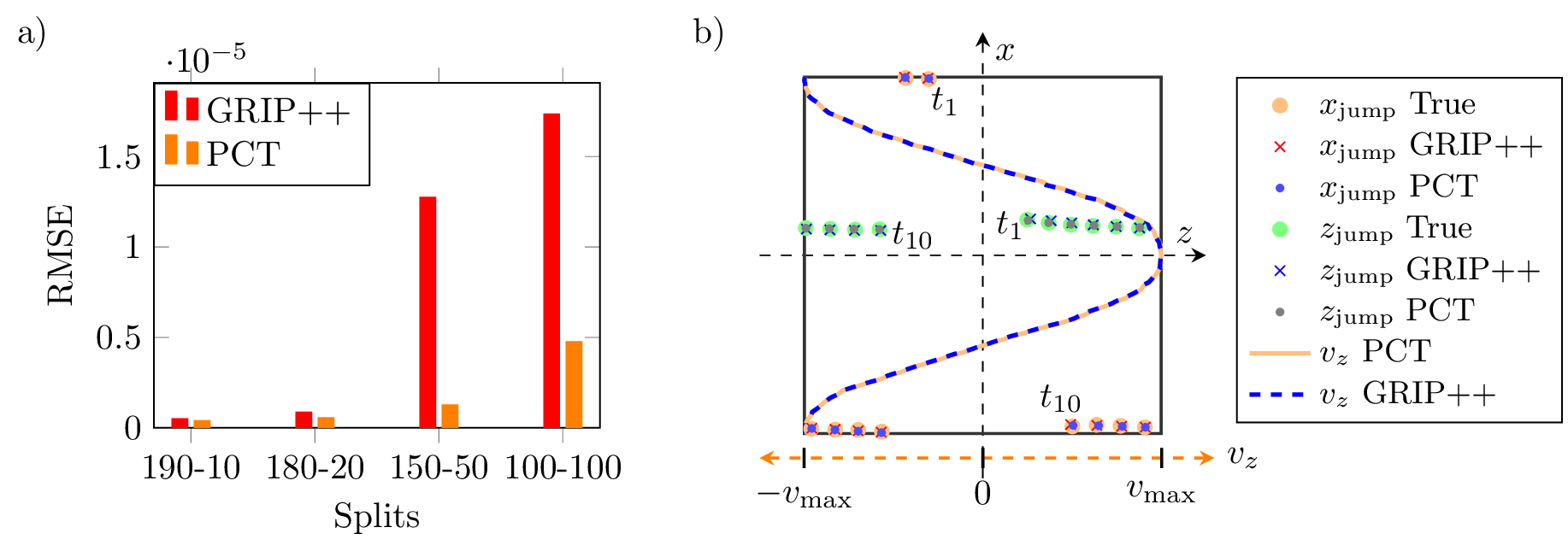}
    \captionsetup{width=\textwidth}
    \caption{a) RMSE comparison between the GRIP++ and PCT methods over various splits of input and output time steps. b) The trajectory of two sample particles in the case of 190-10 split is presented over output time steps. For each particle, the beginning step is denoted by $t_1$ and the end step is denoted by $t_{10}$. The direction of flow of a particle is understood by the shown velocity profile. One particle has jumped over the $x$ and $z$ boundaries and another particle has jumped over the $z$ boundary. In addition, the predicted velocity profile in $z$-direction with respect to the $x-$axis of all the particles has been considered and they are averaged over ten output time steps. This average predicted velocity profile is presented for the two models. The velocity axis is shown by a dashed orange line.}
    \label{fig:5_Periodic_Results}
\end{figure}

We mainly focus our analysis on the RMSE scores as the performance metric and compare the modified GRIP++ and the PCT-based model. In Figure \ref{fig:5_Periodic_Results}a, we show that the PCT-based model performs better than the modified GRIP++ model. As the number of time steps to be predicted increases, the performance difference between the two models gets larger. In addition, for better visualization, we consider the 190-10 split case and show the trajectories of two sample particles exiting and reentering $x$ and $z$ boundaries, see Figure \ref{fig:5_Periodic_Results}b. For this use case, both models were able to match very closely the true particle positions. Also, the velocity profiles of both models are in agreement with the true profile (refer Figure \ref{fig:1_Periodic}). However, the PCT method has a lower overall RMSE score compared to the simply modified GRIP++ model. Note that the VCT method is not employed here for the simple PBCs, as it does not have a discontinuous trajectory as in LEBCs, where the VCT method is particularly useful.

As a supplementary observation, we have discovered that particles may overlap in their predicted trajectories. This phenomenon is illustrated using histograms presented in the Appendix \ref{sec:appendix_periodic}.

\subsection{Evaluation on LEBCs}

\begin{figure}[H]
    \centering
    \includegraphics[width=\textwidth]{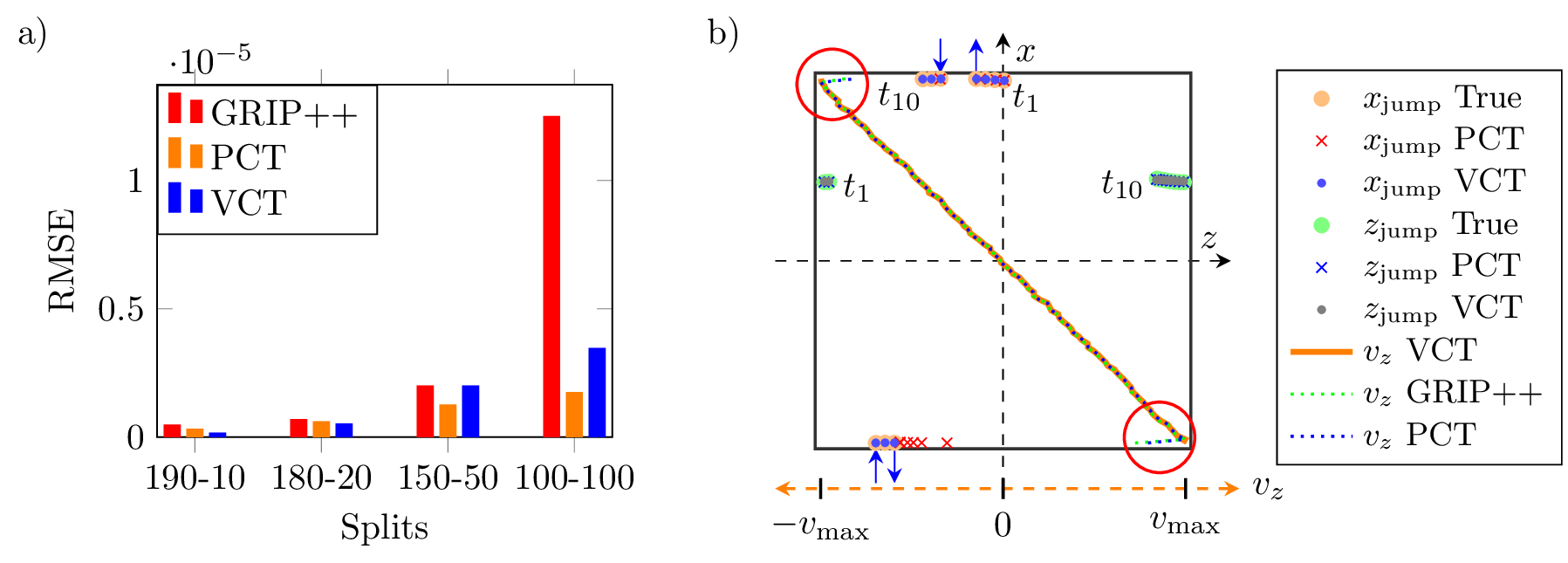}
    \captionsetup{width=\textwidth}
    \caption{a) RMSE comparison among the modified GRIP++ model, the PCT-based, and VCT-based models over various splits of input and output time steps. b) The trajectory of two sample particles in the case of 190-10 split is presented over output time steps. For each particle, the beginning step is denoted by $t_1$ and the end step is denoted by $t_{10}$. The direction of flow of a particle is understood by the shown velocity profile. One particle has jumped over the $x$ boundary and another particle has jumped over the $z$ boundary. The particle exiting and reentering over the $x$ boundary is marked with arrows. In addition, the predicted velocity profile in the $z$ direction with respect to the $x-$axis of all the particles is shown (averaged over ten output time steps). This average predicted velocity profile is presented for the three methods. The velocity axis is shown by a dashed orange line.}
    \label{fig:6_Lees_Edwards_results}
\end{figure}

Handling LEBCs is more difficult than handling periodic ones. This is because the trajectories are discontinuous, even with the introduced data transformation step. In this section, we evaluate the performance of the modified GRIP++ model and the proposed PCT and VCT-based models on the dataset with Lees\texttt{-}Edwards boundary conditions. In Figure \ref{fig:6_Lees_Edwards_results}a, we show how both models, PCT and VCT-based, were able to perform better than the modified GRIP++ model. It can be seen, however, that the VCT-based model is more accurate when making short-term predictions, while the PCT-based model is more accurate for long-term predictions. Consider the velocity profile in Figure \ref{fig:6_Lees_Edwards_results}b, in which focusing on the red circles closer to the top and bottom $x$ boundaries proves that both, the modified GRIP++ model and the PCT-based model, are facing difficulties in predicting velocities correctly at the boundaries. In contrast, the VCT-based model with the help of the introduced feature engineering was able to predict the velocity profile very close to the true profile. In addition, we present the trajectories of two sample particles exiting and reentering the $x$ and $z$-boundaries. For visualization purposes, we compare only the predictions made by the PCT and VCT-based models. The results show that the VCT-based model is more accurate than the PCT-based one when particles are exiting the $x$ boundary.

\begin{figure}[H]
    \centering
    \includegraphics[width=\textwidth]{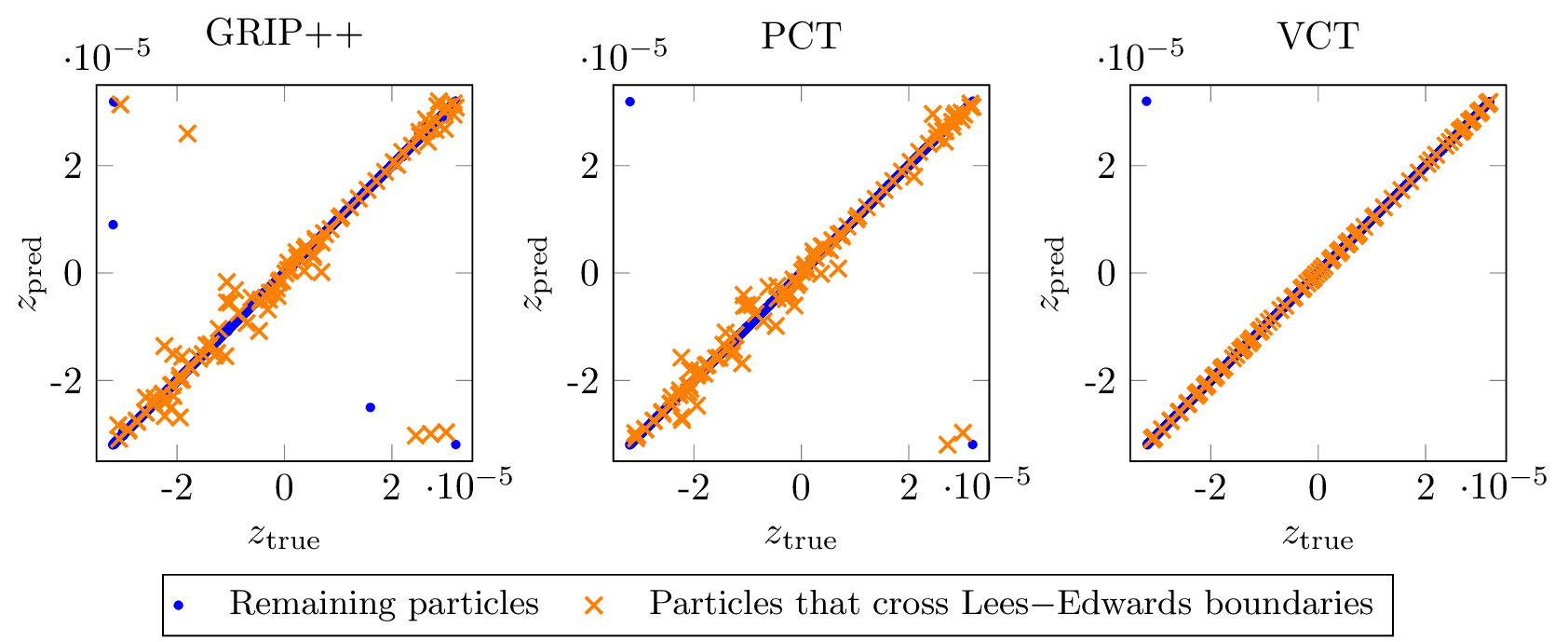}
    \captionsetup{width=\textwidth}
    \caption{Parity plots of the 190-10 split case in which the true vs. predicted $z$ position coordinates are plotted for output time steps. The particles that cross Lees\texttt{-}Edwards boundaries are separated from the remaining particles to show the performance improvement by the VCT method.}
    \label{fig:7_Parity_plots}
\end{figure}

For a further analysis of the results of the VCT-based model, consider the parity plots in Figure \ref{fig:7_Parity_plots} for the 190-10 split case. We separate the particles that cross Lees\texttt{-}Edwards boundaries from the remaining particles to better analyze the prediction behavior of the three applied models. This perspective of the results shows that the particles exiting and reentering $x$ boundary are challenging to learn and require special attention such as feature engineering. Specifically, the VCT-based model is fed with data where we explicitly inform the shifts of the particles that cross Lees\texttt{-}Edwards boundaries. 

\begin{figure}[H]
    \centering
    \includegraphics[width=\textwidth]{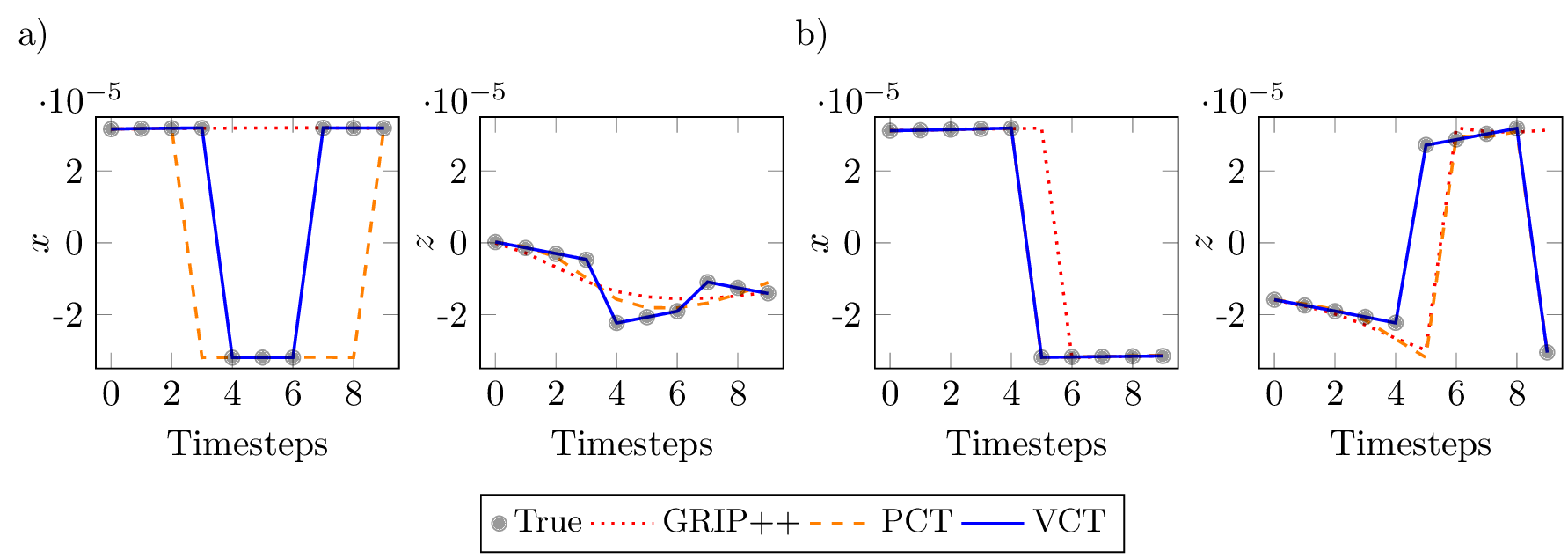}
    \captionsetup{width=\textwidth}
    \caption{Two sample particles crossing the Lees\texttt{-}Edwards boundaries are tracked over output time steps. Trajectory predictions are compared among GRIP++, PCT-based, and VCT-based models.}
    \label{fig:8_Tracking}
\end{figure}

In addition to the above results, we analyze the temporal behavior of two selected particles for the 190-10 use case. Consider Figure \ref{fig:8_Tracking}, where we focus on tracking two particles crossing Lees\texttt{-}Edwards boundary conditions over ten output time steps. For each particle, an illustration of the $x$ and $z$ position coordinates over time steps is presented. The plot shows that the VCT-based model accurately captures the particle trajectory, while the original GRIP++ model and the PCT-based one either predict an averaged behavior (Figure \ref{fig:8_Tracking}a (right)), predict the jump over the boundary too late (Figure \ref{fig:8_Tracking}b or do not capture the trajectory at all (Figure \ref{fig:8_Tracking}a (left)).

Analogous to the results described in Section \ref{sec:results_periodic}, we observed predictions of particles, which are overlapping. These are presented using histograms in Appendix \ref{sec:appendix_LE}.

\section{Discussion}\label{sec:discussion}

The results of our study show that the GRIP++ model originally developed for autonomous driving can be modified to predict particle flow in a representative unit cell. The data transformation step is introduced in this paper in order to get continuous particle trajectories in the case of PBCs. However, the modified GRIP++ model is prone to error accumulation if there is a large number of time steps for which the particle trajectories have to be predicted. This is the case for both our evaluation data sets, cf. Figures \ref{fig:5_Periodic_Results} and \ref{fig:6_Lees_Edwards_results}. Especially, when having a setting with LEBCs, the modified version of GRIP++ has problems to correctly predict the particle trajectory that is crossing such a boundary, see Figure \ref{fig:7_Parity_plots}.

In contrast, the PCT method introduced in this paper improves the prediction performance for all of the analyzed dataset splits. Especially, for a large number of time steps for which trajectories were to be predicted, the PCT-based models outperformed the modified GRIP++ model significantly. This observation holds for both of the evaluation datasets, hence, we assume the model to be less prone to error accumulation. However, when looking at the dataset with LEBCs, still not all of the particles that cross such a boundary are predicted correctly, cf. Figures \ref{fig:7_Parity_plots} and \ref{fig:8_Tracking}. This is caused by the discontinuous trajectories, which are in general challenging to predict.

To tackle this issue, we developed the VCT method along with the feature engineering that differentiates discontinuous data from continuous data. This enables learning the velocities of the particles based on position. By doing so, the VCT-based model is able to track the particles with the LEBCs. Figure \ref{fig:6_Lees_Edwards_results} shows that this approach works if we are looking at the 190-10 and 180-20 splits. However, when there are larger time steps for the trajectories to be predicted, the VCT-based model has difficulties. We assume that this is due to error accumulation, as the VCT-based model is based on predicting velocities rather than positions (like for the PCT method). Nevertheless, for the smaller splits, the VCT-based model reaches the best prediction accuracy for particles that cross Lees\texttt{-}Edwards boundaries, cf. Figures \ref{fig:7_Parity_plots} and \ref{fig:8_Tracking}. 

Before concluding our study, we will discuss the limitations of the presented adaptation and enhancements of the GRIP++ model. Notably, the current implementation does not account for particles of varying sizes and shapes, which can significantly influence particle flow. However, this information can be integrated into the approach by adding additional parameters for each particle's location. Specifically, the following can be included: (i) the radius of each particle to define its size, and (ii) additional parameters to characterize the particle's shape and orientation. While incorporating the radius involves minimal effort, adding shape and orientation parameters can significantly increase the problem's dimensionality, leading to higher computational costs and greater complexity in the learning task. Furthermore, predicting macroscopic responses such as stresses presents a unique challenge due to their inherent instability; stresses tend to spike when particles make contact and then rapidly decrease back to zero. The prediction of stress trajectories, however, was not in the scope of our study.

Furthermore, we encountered that all of the approaches described above the prediction of particles that are overlapping. From a physical standpoint, in our use case, the overlapping of rigid particles is not possible. This occurs because the model, when predicting the position of particles for the next time step, cannot determine if a particle already occupies a specific location. However, there is no mechanism to prevent the prediction models from that behavior. This is also true for the original GRIP++ model for vehicle trajectory prediction. 

\section{Summary and Outlook}\label{sec:conclusion}

In the present paper, we enhanced the graph-based interaction-aware trajectory prediction model GRIP++, originally developed for the transportation domain, to predict particle trajectories in a discrete element simulation of a representative unit cell. A modification of the original model to handle 3D use cases was conduced to enable the model to predict the particle motion to some extent. In addition, we introduced two enhanced models that outperformed the original model as adapted to the 3D case. Furthermore, we introduced a data transformation step to incorporate periodic boundary conditions into the models. The results of our study are promising, indicating that our enhanced models can effectively capture the complex dynamics of particle interactions and provide accurate trajectory predictions. Additionally, our findings suggest that it is generally feasible to adapt models from the transportation domain for use in particle simulation. This opens the possibility for further exploration of other models that may also be transferable to this domain.

Nevertheless, the models studied in this work can act as a starting point to incorporate machine learning into discrete element simulations for forecasting particle trajectories. Moving forward, the next steps are to demonstrate the model's functionality in the example of simulating real-world particle flows for example in a manufacturing process. Additionally, future work includes developing a model that is not limited to representative unit cell environments. We also aim to develop particle trajectory prediction models that incorporate lubrication, magnetic, and other physical forces that influence particle flows.

\section*{Data availability statement}
Data and code will be made available upon reasonable request.

\section*{Competing Interests}
The authors have no competing interests to declare that are relevant to the content of this article.

\section*{Funding sources}
This research did not receive any specific grant from funding agencies in the public, commercial, or not-for-profit sectors.

\section*{Author contribution}
AS was responsible for the development of the code for the modified GRIP++ model, including all enhancements presented in the paper, and applied the model to the given use cases while also conducting evaluations. AS worked closely with LM in writing the manuscript. LM provided valuable insights in editing the draft and supervised the project. The idea to extend the GRIP++ model to the 3D case and apply it to problems outside the transportation domain was conceived by PR. CB provided valuable insights and guidance from the discrete element simulation domain. Furthermore, CB provided the use cases and the datasets. All authors participated in the discussion of the results and in reviewing the manuscript.

\bibliographystyle{ieeetr}
\bibliography{main}

% Appendices
\clearpage % or \newpage
\appendix

\section{Particles Overlapping}
\label{sec:appendix}

\subsection{Evaluation on Simple Periodic Boundary Conditions}
\label{sec:appendix_periodic}

\begin{figure}[H]
    \centering
    \includegraphics[width=0.73\textwidth]{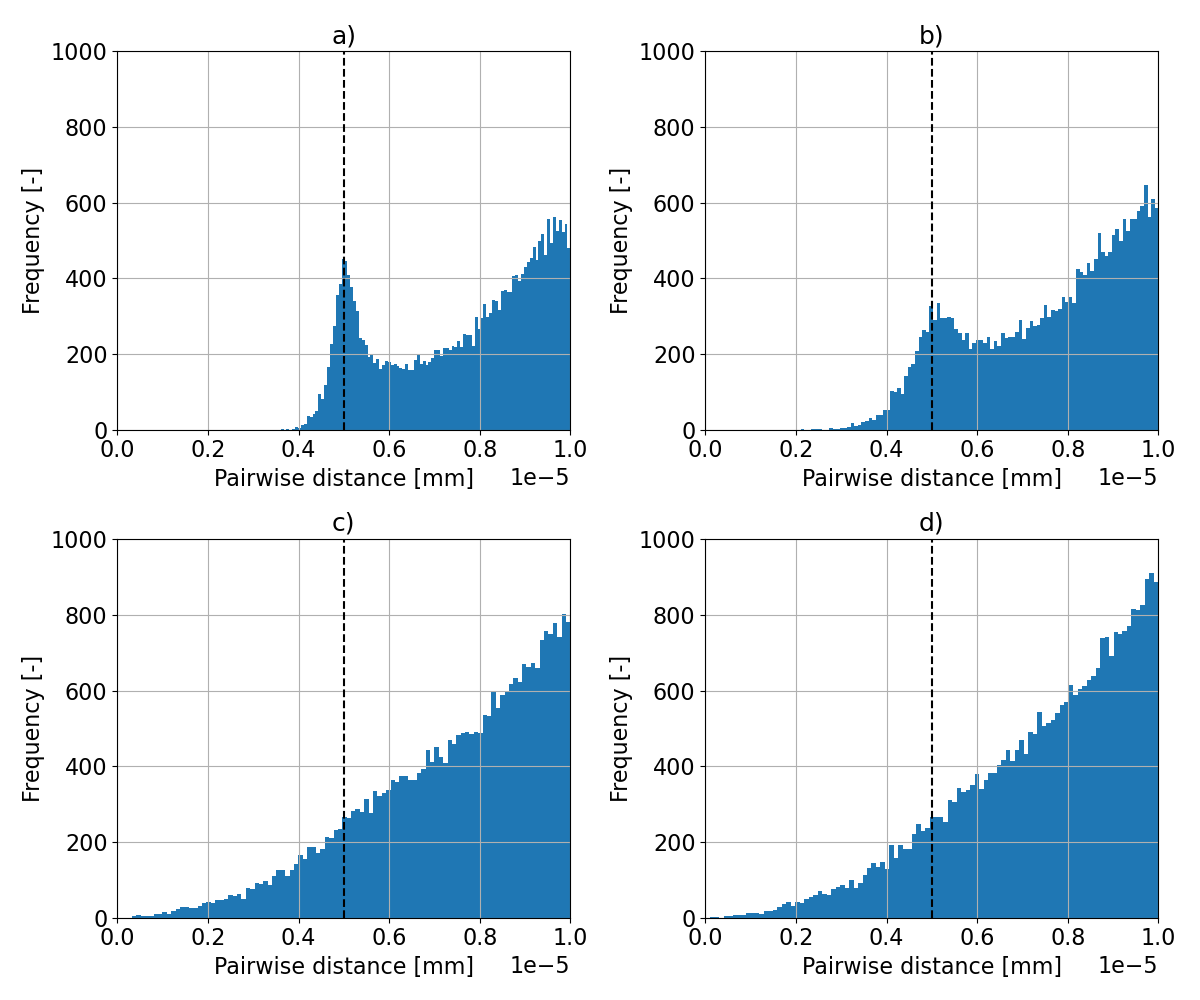}
    \captionsetup{width=\textwidth}
    \caption{Particles overlapping as predicted by the enhanced GRIP++ model. Depicted is the pairwise Euclidean distance between the particles predicted at the last time step of splits (a) 190-10, (b) 180-20, (c) 150-50, (d) 100-100. The dashed black line marks the particle diameter. The plot is truncated at a pairwise distance of $10^{-5}$.}
    \label{fig:appendix_periodic_GRIP++}
\end{figure}

\begin{figure}[H]
    \centering
    \includegraphics[width=0.73\textwidth]{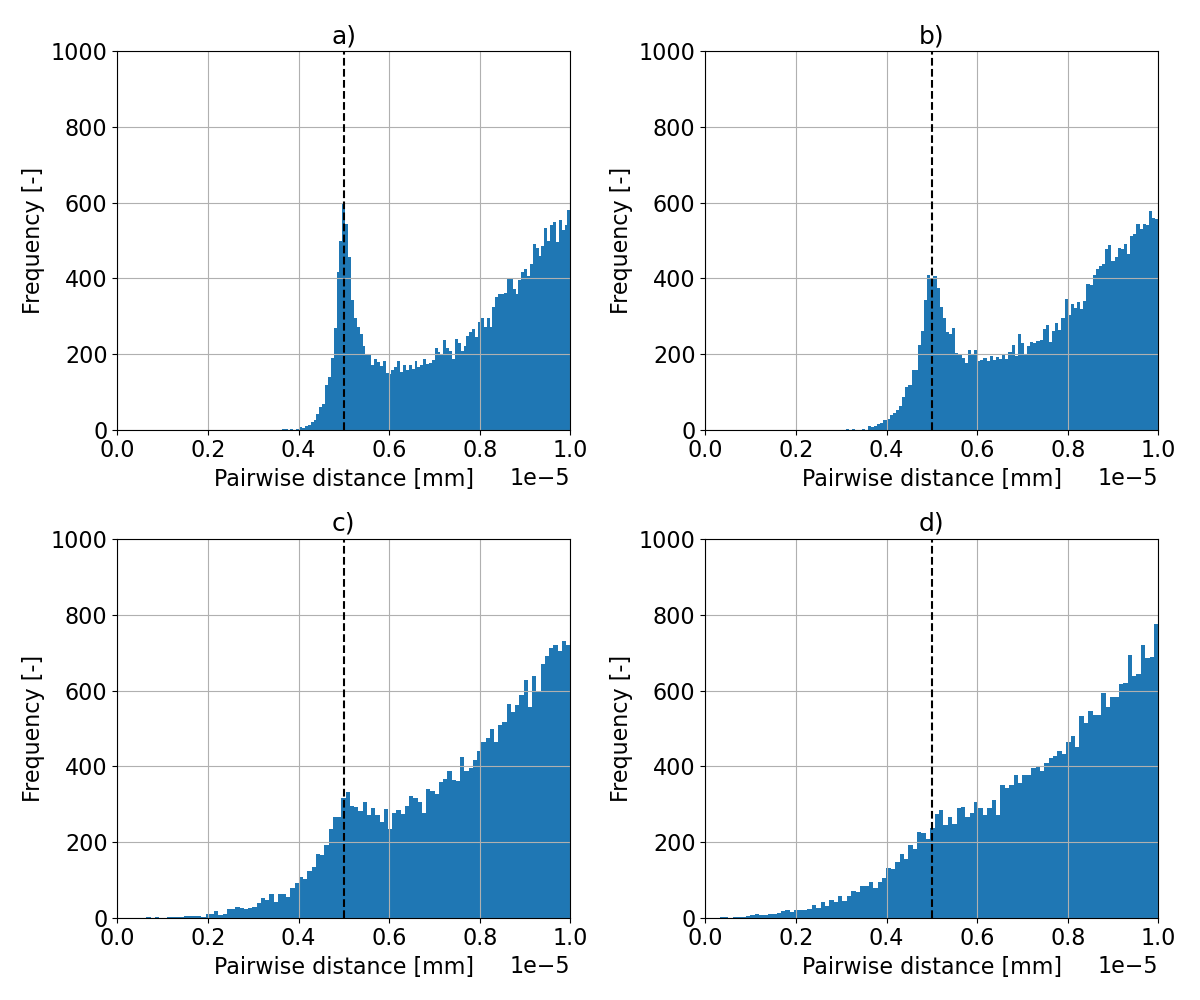}
    \captionsetup{width=\textwidth}
    \caption{Particles overlapping as predicted by the PCT-based model. Depicted is the pairwise Euclidean distance between the particles predicted at the last time step of splits (a) 190-10, (b) 180-20, (c) 150-50, (d) 100-100. The dashed black line marks the particle diameter. The plot is truncated at a pairwise distance of $10^{-5}$.}
    \label{fig:appendix_periodic_PCT}
\end{figure}

To quantify the extent of particle overlap in predicted locations in our first example, we calculated the pairwise Euclidean distances between all particles at the final time step across all evaluation splits. The resulting distributions are shown in Figures \ref{fig:appendix_periodic_GRIP++} and \ref{fig:appendix_periodic_PCT} in the form of histograms. The dashed black line in each figure represents the particle diameter, with values below this threshold denoting instances of particle overlap.

\subsection{Evaluation on Lees\texttt{-}Edwards Boundary Conditions}
\label{sec:appendix_LE}

\begin{figure}[H]
    \centering
    \includegraphics[width=0.73\textwidth]{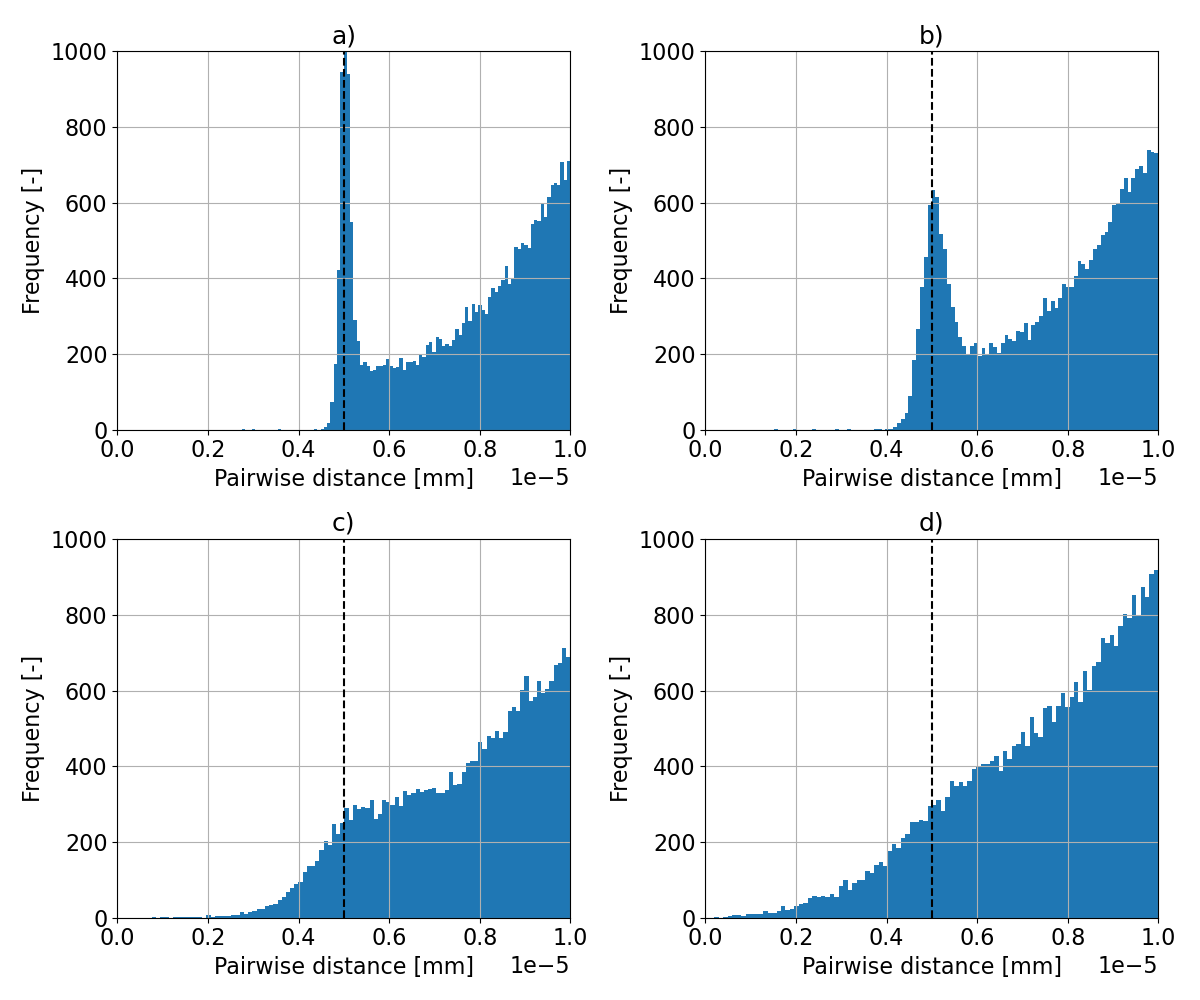}
    \captionsetup{width=\textwidth}
    \caption{Particles overlapping as predicted by the enhanced GRIP++ model. Depicted is the pairwise Euclidean distance between the particles predicted at the last time step of splits (a) 190-10, (b) 180-20, (c) 150-50, (d) 100-100. The dashed black line marks the particle diameter. The plot is truncated at a pairwise distance of $10^{-5}$.}
    \label{fig:appendix_LE_GRIP++}
\end{figure}

\begin{figure}[H]
    \centering
    \includegraphics[width=0.73\textwidth]{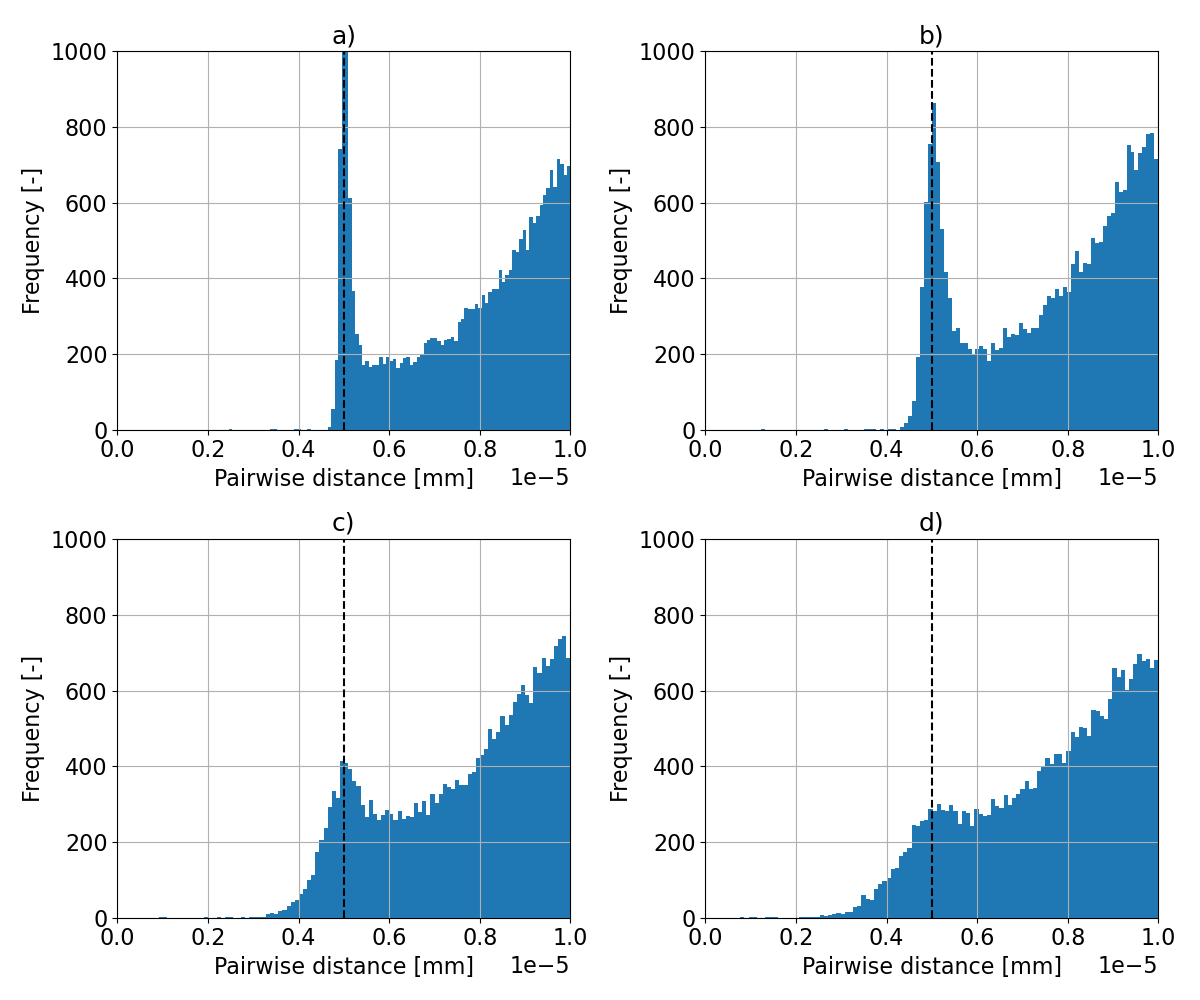}
    \captionsetup{width=\textwidth}
    \caption{Particles overlapping as predicted by the PCT-based model. Depicted is the pairwise Euclidean distance between the particles predicted at the last time step of splits (a) 190-10, (b) 180-20, (c) 150-50, (d) 100-100. The dashed black line marks the particle diameter. The plot is truncated at a pairwise distance of $10^{-5}$.}
    \label{fig:appendix_LE_PCT}
\end{figure}

\begin{figure}[H]
    \centering
    \includegraphics[width=0.73\textwidth]{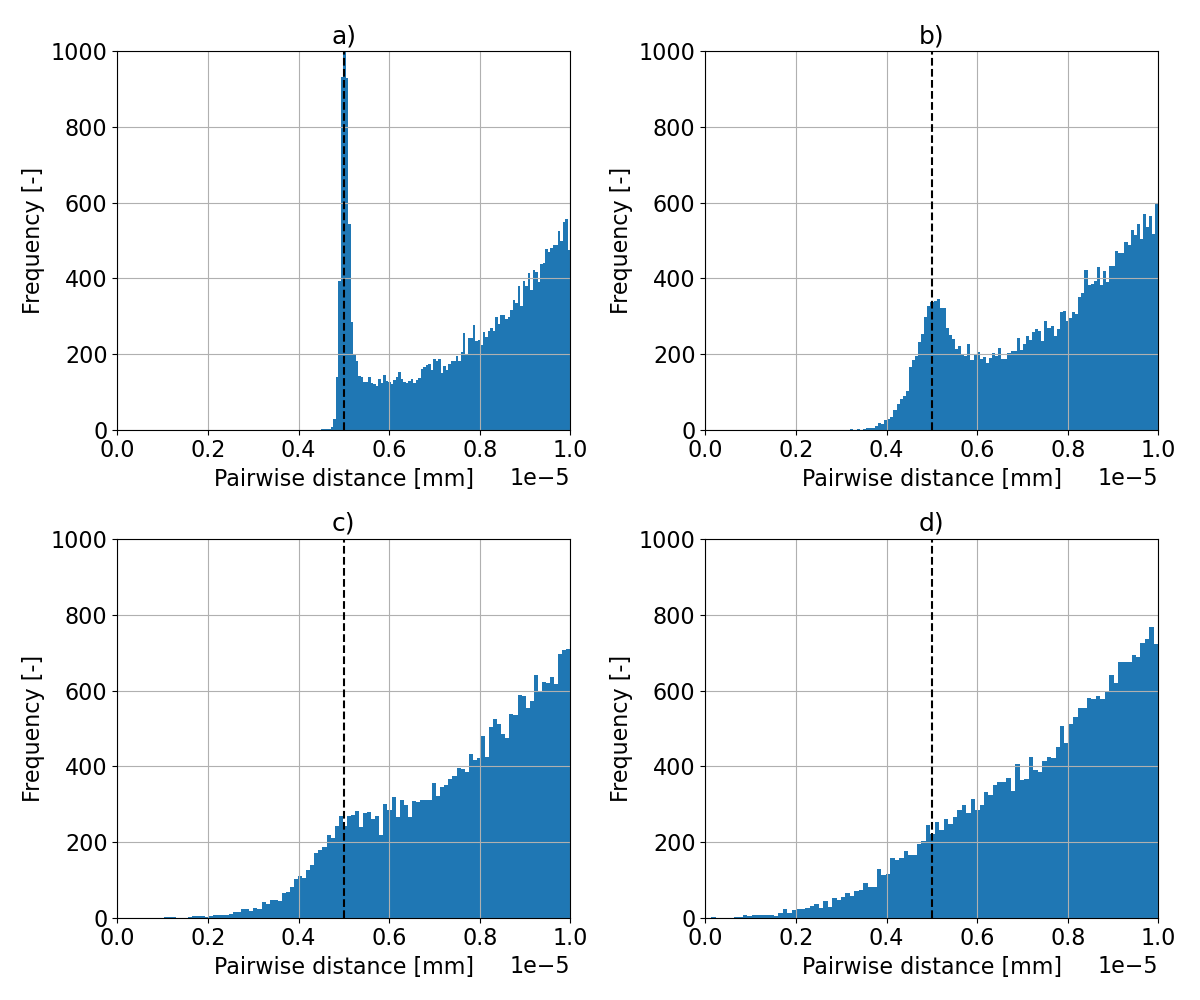}
    \captionsetup{width=\textwidth}
    \caption{Particles overlapping as predicted by the VCT-based model. Depicted is the pairwise Euclidean distance between the particles predicted at the last time step of splits (a) 190-10, (b) 180-20, (c) 150-50, (d) 100-100. The dashed black line marks the particle diameter. The plot is truncated at a pairwise distance of $10^{-5}$.}
    \label{fig:appendix_LE_VCT}
\end{figure}

To assess the extent of particle overlap in predicted locations in our second example, we computed the pairwise Euclidean distances between all particles at the final time step across all evaluation splits. The resulting distribution is presented as a histogram in Figures \ref{fig:appendix_LE_GRIP++}, \ref{fig:appendix_LE_PCT}, and \ref{fig:appendix_LE_VCT}. The dashed black line in each figure represents the particle diameter, with values below this threshold indicating instances of particle overlap.

\end{document}